\documentclass[]{article}

\def\ben{\begin{equation}}
\def\een{\end{equation}}
\def\bea{\begin{eqnarray}}
\def\eea{\end{eqnarray}}
\input amssym.def
\input amssym.tex
\begin{document}

\hfuzz=100pt
\title{Pulse Propagation in Born-Infeld Theory, the World-Volume 
Equivalence
Principle and the Hagedorn-like Equation of State of the Chaplygin Gas}
\author{G. W. Gibbons
\\
D.A.M.T.P.,
\\ Cambridge University, 
\\ Wilberforce Road,
\\ Cambridge CB3 0WA,
 \\ U.K.}
\maketitle

\begin{abstract} 
A recently proposed world-volume equivalence principal
involving the Boillat, as opposed to the Einstein, metric
is examined in the context of some colliding wave solutions
of the Born-Infeld equations for which two
plane polarized pulses  pass through one another without distortion.
They suffer a delay with respect to the usual Einstein
metric
but not with respect to the Boillat metric.
Both metrics are flat in this case, and  the closed string and open string
causal structures
are interchanged by the Legendre transformation that is used for
solving
the associated Monge-Amp\`ere equation. In 1+1 dimensions
the equations are known to be equivalent to the vorticity
free motion of a Chaplygin gas.
The latter is shown to be described by the  
scalar Born-Infeld equation in all dimensions and  it is pointed 
out that the equation
of state is Hagedorn-like: there  is
an upper bound to the pressure
and temperature.

\end{abstract}
\section{Introduction}
In two recent papers \cite{GH, GW} it has been argued that
the propagation of  excitations
around D-branes, 
and also those around the M5-brane,  satisfy a 
a  generalization of the usual
Einstein  Equivalence Principal. One finds that   
the equations of motion  
are universally
governed by a co- metric $C^{\mu \nu}$, called there the Boillat metric,
which, in the presence of vector fields or tensor fields
differs from the pullback $g_{\mu \nu}$ of the usual Einstein metric
in such a way that, using just world volume fields, the Einstein metric
$g_{\mu \nu}$ is effectively unobservable. In string theory the
Einstein metric is the closed string metric whle the Boillat metric
is conformal to the open string metric.
In \cite{GW}
analogy was drawn with treatments of general relativity
in which one introduces an auxiliary flat metric $\eta_{\mu \nu}$.
Because of the usual Einstein Equivalence Principal any such
flat metric is in effect unobservable.  The characteristic co-metric of 
any set of equations of motion is determined only up to to a Weyl
re-scaling. In the case of D-branes 
one might for example  use instead the open string co-metric $G^{\mu \nu}$.
This is conformal to the Boillat metric.
The choice of the particular conformal
representative  $C^{\mu \nu}$ may be  motivated by at least two arguments.
\begin{itemize}
\item In the case of the D3-brane, $C^{\mu \nu}$, unlike $G^{\mu \nu}$ is
invariant under S-duality, i.e. under electric-magnetic duality rotations.
\item In both the case of D-branes and for the M5-brane, the following
relation holds between the energy momentum tensor $T^{\mu \nu}$
of the Born-Infeld field or of the closed 3-form field
\ben
T^{\mu \nu}= C^{\mu \nu}-g^{\mu \nu} \label{Hooke}. 
\een
For reasons explained in \cite{GW} (\ref{Hooke}) may be referred
to as  Hooke's Law. In the Born-Infeld case one has the further
remarkable determinental identity
\ben
\det (C^{\mu \nu }) = \det(g^{\mu \nu}), \label {det1} 
\een
which implies that
\ben
\det \bigl (T^\mu _\nu + \delta ^\mu_\nu \bigr )=1. \label{det2}. 
\een
\end{itemize}

The aim of the present paper is to explore this putative 
World-Volume Equivalence Principle in greater detail in the
simplified situation of $1+1$ dimensions. We shall rely heavily
on a  recent paper \cite{IP} on the propagation of pulses
in non-linear electrodynamics (see also earlier work of
Barbashov and Chernikov)   and we shall make contact with some
older work of Schr\"odinger \cite{S} on Born-Infeld theory
as a theory of non-linear optics and  work on
the relation between scalar Born-Infeld, minimal surfaces, strings
and the Monge-Amp\`ere equation. In fact the remarks in the 
discussion section of \cite{IP} in their $1+1$ dimensional
setting have much in common with the suggestions made in \cite{GH, GW}.

In attempting to generalize these results to higher dimensions,
we are led to consider the scalar Born-Infeld equation in higher
dimensions.
This is found to be equivalent to a Chaplygin gas. It has a
Hagedorn-like  equation of state with an upper bound for the
pressure and temperature. This may be relevant to recent 
attempts to use fluid dynamic ideas to discuss the strong coupling
limit open strings \cite{GHY, AS}.

\section{ Hooke's Law, the Monge-Amp\`ere Equation and Pulse 
interactions}

The striking determinental identity (\ref{det1})
has an immediate  application to the propagation of pulses
in Born-Infeld theory.

In flat two dimensional spacetime, the conservation law for 
the stress tensor implies that it is given
by a single free function, the  Airy stress function 
$\psi$, \cite{A} such that 
\ben
T_{tt}= \psi_{zz}, \quad T_{zz} = \psi_{tt}, \quad T_{tz}=\psi_{zt}.  
\een

Now for a {\sl single}
 polarization state for which the only non-vanishing
component of the electric field is $E_x$ , the determinental
identity (\ref{det1}) still holds when restricted to just
the  propagation directions $(t,z)$ even  though
the two form $F_{\mu\nu}$ is non-vanishing
in the $(t,x)$ and $(z,x)$ directions.

Written in terms of the Airy stress function, 
the determinental identity (\ref{det1}) then becomes 
a special case of the  Monge-Amp\`ere equation
\ben
\psi _{zz} \psi _{tt}- \psi ^2 _{zt}= \psi_{zz}-\psi_{tt}
\een
This can be solved exactly \cite{IP} using the hodographic
method, that is performing a Legendre transform
under which it becomes D'Alembert's equation with respect to 
a new set of variables $T$ and $Z$. 
  
The solution given in \cite{IP} gives
\ben
T^{tt}={A+B+2AB \over 1-AB},
\een
\ben
T^{zz}={A+B-2AB \over 1-AB},
\een
\ben
T^{zt}={B-A \over 1-AB },
\een
where $A=A(T+Z)$ and $B=B(Z-T)$ are arbitrary functions of their arguments.
The relation between the new coordinates $(T,Z)$ and usual coordinates
$(t,z)$ is most conveniently expressed using null coordinates.
Let $v=t+z, u=t-z, \xi= T+Z, \eta=Z-T$. The asymmetrical
definition of $\eta$ is so as to agree with \cite{IP}, One has 
\ben
dv=d\xi-Bd \eta, \qquad du= -d\eta + A d\xi.
\een
Thus
\ben
(1-AB) d \eta=Adv-du, \qquad (1-AB) d \xi=  dv-Bdu.
\een
one checks that   
\ben
dT^2-dZ^2= dt^2(1-A-B+AB) -dz^2(1+ A+B +AB)-2dtdz(A-B) =
C^{-1}_{\mu \nu} dx^\mu dx^nu,
\een
where
\ben
C^{\mu \nu}= \eta^{\mu \nu} + T^{\mu \nu}.
\een
Thus we see that the Legendre transformation
to the new coordinates $(T,Z)$ used to solve
the Monge-Amp\`ere equation in effect passes to flat
inertial coordinates with respect to the 
Boillat metric. It should be noted that one does not expect
the Boillat metric to be flat in general in higher dimensions.

The general solution consists of two pulses, one right-moving and one left moving
which pass through one another without distortion. 
In terms of the usual coordinates $(t,z)$
the two pulses experience a {\sl delay} That is measured with respect to the closed string metric. However
With respect to the Boillat coordinates, that is measured with
respect to the Boillat metric, there is no delay.

\section{Legendre Duality} 

As noted above, the two causal structures are permuted by 
Legendre duality. It is of interest to examine this more fully.
One thinks of $(t,z)$ as coordinates for ${\Bbb E}^{1,1}$ and 
$(T,Z)$ as coordinates for the dual of the dual
momentum space $ \Bigl ({ {\Bbb E}^{1,1} }^\star  \Bigr )^\star$.
Because of the Lorentzian metric one includes an
additional  minus sign
in the Legendre transformation. Put another way, one composes the
standard Legendre map between ${\Bbb E}^{1,1}$ and ${{\Bbb E}^{1,1}}^\star$
with the musical isomorphism $\sharp$ (``index raising ")
between ${{\Bbb E}^{1,1}}^\star$
and ${\Bbb E}^{1,1}$ to get a map between 
${\Bbb E}^{1,1}$ and ${\Bbb E}^{1,1}$. 
The two coordinate charts $(tz)$ and $(T,Z)$ give rise to what Maxwell 
called, in the context of statics, reciprocal diagrams, one
of position and one of stress \cite{Max}. 
The map between the 
two diagrams is also
referred to as the hodographic transformation.

Explicitly
\ben
T= {\partial \psi \over \partial t} \qquad Z=- 
{\partial \psi \over \partial z}.
\een
define the Legendre conjugate  function $\phi(T,Z)$ by
\ben
\phi =Tt-Zz- \psi,
\een
Thus
\ben
t={\partial \phi \over \partial T} \qquad z=-{\partial \phi \over \partial Z}.
\een
It now follows that  
Hessian of $\psi$ composed with index raising is the inverse
of the Hessian of  $\phi$. In two dimensions,
up to a factor of the inverse of the determinant, 
the inverse of a matrix  is a linear function of the elements of a matrix
and this allows us to 
linearize the Monge-Amp\`ere equation, which becomes in terms of
$\phi$
\ben
\phi_{TT}-\phi _{ZZ}=1.
\een
The general solution is  
\ben
\phi ={ 1 \over 2} (T^2 -Z^2) + f_L(T-Z) + f_R(T+Z),
\een
where the arbitrary right-moving $F_L(T-Z)$ and left-moving $f_R(T+Z)
$ waves
are related 
in a simple way to the functions $A(T+Z)$ and $B(Z-T)$ introduced   
in \cite{IP}.

Note that if we set
\ben
\psi = u + { 1\over 2} (t^2-z^2),
\een
then $u$ satisfies a more familiar form of the Monge-Amp\`ere 
equation
\ben
u_{tt} u_{zz} -u^2_{tz}= -1.
\een

This equation is Legendre self-dual, so using the hodograph method 
directly will not lead to a solution.

\section{Additional Polarization States}

The case considered in \cite{IP} was of a single polarization state.
Now it if one includes all possible polarization states
in ${\Bbb E}^{n-1,1}$, that is, 
if we assume that
$A^\mu=(0,0, A^i (t,z) )$, $i=2,3,\dots ,n-2$  
then \cite{G} the  Born-Infeld action reduces
to a gauge-fixed  Dirac-Nambu-Goto action for a string
\ben
{ 1\over 2} \int  \Bigl (1- \sqrt {1- (\partial _t A^i)^2 +(\partial _t A^i)^2 } \Bigr ) dt dz
\label{Nambu} \een

One has
\ben
T_{tt}+1 = { 1+ (A^i_z)^2  \over 
\sqrt {1- (\partial _t A^i)^2 +(\partial _z A^i)^2 } }
\een
\ben
T_{zz} -1 = 
{ 1-(A^i_t)^2  \over \sqrt {1- (\partial _t A^i)^2 +(\partial _z A^i)^2 } }
\een
\ben
T_{tz}= { A^i_zA^i_t  \over \sqrt {1- (\partial _t A^i)^2 +(\partial _t A^i)^2 }  }
\een

For one polarization state, the determinental identity holds,
but for more than one, it does not unless. 
                                         
\ben
(A^i_tA^i_z)^2= (A^k_t)^2(A^j_z)^2. 
\een

The case of just one {\sl plane} polarization state is
called the (1+1)-dimensional Born-Infeld equation
and is related to other completely integrable systems. 
For $n=4$ and the case of {\sl circular polarization} Schr\"odinger \cite{S}
showed that one could superpose solutions. 
Because of the obscurity of the reference it may be helpful
to desribe his method. One defines
\ben
{\bf F}={\bf B}-i{\bf D}, \qquad {\bf G}={\bf E}+i {\bf H}.
\een
One needs to solve the {\sl linear} equations
\ben
{\rm div} \\\ {\bf F}=0, \qquad {\rm curl} \\\ {\bf G}+ {\partial {\bf F}
\over \partial t}=0, 
\een
subject to the {\sl non-linear} constitutive relation
\ben
{\overline {\bf G}}= { 2 {\bf F} ({\bf F}.{\bf G})- ( {\bf F}^2 -{\bf
G}^2 ) {\bf G} \over ({\bf F} .{\bf G})^2 }.
\een

Schr\"odinger showed that exact solutions exist of the form
\ben
{ \bf F}=  C_1{\bf a}_1 e^{ i({\bf k}_1.{\bf x}-\omega_1 t)} +
 C_2{\bf a}_2 e^{i({\bf k}_2.{\bf x}-\omega_2 t)}
\een 

\ben
{ \bf G}=i A_1  C_1{\bf a}_1 e^{i({\bf k}_1.{\bf x}-\omega_1 t)} +iA_2 C_2{\bf a}_2 e^{i({\bf k}_2.{\bf x}-\omega_2 t)}
\een 
 for appropriate choices of the constants. By Lorentz-invariance
the two waves may be taken to  move in opposite 
directions and Scr\"odinger  showed 
that they do so with speeds which do not exceed that of light.

In \cite{G, GH} it was suggested that this is a general
feature for Born-Infeld. The argument was that the action
(\ref{Nambu}) is the gauge fixed form of the action
\ben
 - \int \sqrt{ \det (-\eta_{\mu \nu} x^\mu_a x^\nu_b )}dtdz
\een
in Monge (often inappropriately called static) gauge.

Now in conformal gauge:
\ben
\eta_{\mu \nu} x^\mu _t x^\mu _t=\eta_{\mu \nu} x^\mu _z x^\mu _z
\qquad \eta_{\mu \nu} x^\mu _t x^\mu _z=0,
\een
the solutions are the sums of left and right movers, i.e. of functions
of $t-z$ or $t+z$. Thus pulses should pass through one another without
deformation.

\section{Higher Dimensions}

The reduction of the Born-Infeld equation to the Monge-Amp\`ere
equation 
using the Airy stress function in two dimensions suggests looking at
the  analogue in higher dimensions.  Maxwell \cite{Max} provided
a generalization to ${\Bbb E}^3$  which in turn is easily generalized to 
${\Bbb E}^n$ and hence, with appropriate adjustment of signs,
to Minkowski space ${\Bbb E}^{n-1,1}$. In three dimensions one defines
three functions $A,B,C$  by 

\ben T_{xy}=\partial_x \partial _y C \qquad {\rm and 
\\\ cyclically}. 
\een
One then finds that the conservation laws imply, up to
trivial redefinitions that 
\ben
T_{xx}=-\partial_y \partial_y C -\partial _z \partial _z B
\qquad {\rm and \\\ cyclically}  
\een
In $n$ dimensions one defines ${ 1\over 2} n(n-1)$ functions
using the off-diagonal components of the stress tensor
and uses the $n$ conservation equations to deduce the remaining
$n$ diagonal components.  

However, by contrast with the case of two dimensions, imposing 
a vanishing trace condition (which in two dimensions lea
The underlying problem is that in general one has two many functions
and thus no obvious  Legendre  transformation.
Laplace's equation) or the determinental identity (which in two dimensions
leads to the Monge-Amp\`ere
 equation) does not seem to lead to
recognizable integrable equations in general.  In special cases,
for example if $A=B=C$, some simplifications do result. 
A detailed consideration will be  deferred until a later date.

\section{The Higher-dimensional scalar Born-Infeld equation
and the Chaplygin Gas}

This comes from the Lagrangian
\ben
 \int  \Bigl ( 1-\sqrt{ 1- \eta^{\mu \nu} \partial_\mu A \partial _\nu A
} \Bigr ) d^nx.
\een
One may think of $A$ as the transverse coordinate of a domain wall
or $(n-1)$-brane.  
The energy momentum tensor satisfies Hooke's Law
\ben
T^{\mu \nu}+ \eta^{\mu \nu}= C^{\mu \nu},
\een
where the equations of motion take the form
\ben
C^{\mu \nu} \partial_\mu \partial _\nu \phi=0.
\een

The characteristic co-metric $C^{\mu\nu}$ is
essentially the inverse of  induced metric on the brane

\ben
C^{\mu \nu}= \sqrt { 1- \eta^{\mu \nu} \partial_\mu A \partial _\nu A}
 \Bigl (
g^{\mu \nu} + {\partial ^\mu \phi \partial ^\nu \phi \over  1- \eta^{\mu \nu} \partial_\mu A \partial _\nu A }  \Bigl ).  
\een
 
One finds that 

\ben
\det ( T^{\mu \nu}  +\eta^{\mu \nu} ) =  
(1- \eta^{\mu \nu} \partial_\mu A \partial _\nu A ) ^{ n-2 \over 2}.
\een

As with all Lagrangians involving only the partial derivative
of a single scalar, the system is equivalent,
at least for spacelike level sets of the function $A$,
to a vorticity-free
perfect fluid with 4-velocity
\ben
U^{\mu}= {\partial ^\mu  A \over \sqrt{\eta^{\sigma \lambda} 
\partial_\sigma A \partial _\lambda A } }.
\een

The pressure $P$ and energy density $\rho$ satisfy
the equation of state
\ben
(\rho +1) (P-1)=-1 \label{state}.
\een
If $n=2$ the equation of state coincides with the determinental identity.
This type of gas is called a Chaplygin gas and is known to give an
integral system in 1+1 dimensions \cite{N}.
Note that, when defining the energy momentum tensor, there is the freedom
to add a cosmological term. This shifts $\rho$ and $P$ by constants
so that
\ben
(\rho + \lambda) (P-\lambda ) =-\lambda ^2. 
\een

I have fixed this freedom by demanding that $\rho$ and $P$ vanish if 
the scalar field $A$ is constant. If instead I had chosen 
$\lambda =0$ I would have obtained $P=-{ 1\over \rho}$
which is also frequently
referred as a Chaplygin
 equation of state. Because now the pressure is negative,
this has been invoked  to account for recent super-novae evaluations
of the cosmological constant. In effect the Chaplygin gas is a 
form of what has been called "k-essence``.  Form the standpoint adopted 
here, the negative pressure is a result of ``tuning'' 
the arbitrary constant.

The sound speed $c_s$ follows 
from the equation of state
\ben
p= {\rho \over \rho +1}
\een
and Newton's formula (which is equivalent to finding the 
the characteristic cone give by $C^{\mu \nu}$ ) 
\ben
c_s = \sqrt {\partial P \over \partial \rho }. 
\een
One has
\ben
c_s = { 1\over \rho +1}.
\een
 
The sound speed never exceeds that of light, 
equals that of light at low density
(so-called stiff matter) and goes to zero at large density.
This, and the  upper bound for the pressure, is reminiscent
of the Hagedorn behaviour of dual resonance models. To investigate further,
and in particular to see whether
 there is an upper bound for the temperature,
one notes that one may calculate the entropy density $s$
and the temperature $T$ for any vorticity-free perfect fluid 
with velocity potential $A$.  One easily checks that the following
general formulae hold
\ben
T^2= \eta ^{\mu \nu} \partial_\mu   A \partial_\nu A,
\een 
\ben
s^2= J_ \mu J ^\mu,
\een
where the entropy current
\ben
J^\mu =s U^\mu,
\een
is conserved by virtue of the Euler-Lagrange equations. Further
\ben
P=L \qquad \rho+P=Ts.
\een
Thermodynamically, temperature $T$ and entropy density $s$ are
conjugate variables and the pressure $P=P(T)$
and energy density $\rho=\rho(s)$ are related by a 
Legendre transformation
\ben
T={\partial P \over \partial \rho}, \qquad \rho = {\partial \rho \over 
\partial s}. 
\een

For the Born-Infeld/Chaplygin case
\ben
P= 1-\sqrt{1-T^2}, \qquad \rho = \sqrt{1+s^2} -1
\een
\ben
s= {T \over \sqrt {1-T^2} }, \qquad T = {s\over \sqrt{ 1+s^2} },
\een
\ben
\rho = { 1\over \sqrt {1-T^2}} -1, \qquad P= 1- {1 \over \sqrt {1+s^2} }. 
\een

Thus indeed one has an upper bound for the temperature 
just like the Hagedorn temperature associated to the the states
of perturbative string theory. However one may check
that the detailed equation of state in the vicinity of the 
Hagedorn temperature  differs form that given by 
perturbative string theory.

\section{Acknowledgements}

I would like to thank Carlos Herdeiro and Peter West for
helpful discussions on the material of this paper.
I have also benefitted from discussions in the past
with Yuval Nutku on the Monge-Amp\`ere equation.

\end{document}